\documentclass[graybox]{svmult}

\usepackage{mathptmx}       % selects Times Roman as basic font
\usepackage{helvet}         % selects Helvetica as sans-serif font
\usepackage{courier}        % selects Courier as typewriter font
\usepackage{type1cm}        % activate if the above 3 fonts are
\usepackage{makeidx}         % allows index generation
\usepackage{graphicx}        % standard LaTeX graphics tool
\usepackage{multicol}        % used for the two-column index
\usepackage[bottom]{footmisc}% places footnotes at page bottom

\makeindex             % used for the subject index
\begin{document}

\title*{Chemical Evolution of M31}
\author{Francesca Matteucci and Emanuele Spitoni}
\institute{Francesca Matteucci \at Department of Physics, Asctronomy Section, 
Trieste University, Via G.B. Tiepolo, 11, 34131 Trieste \email{matteucci@oats.inaf.it}
\and Emanuele Spitoni \at Department of Physics, Asctronomy Section, Trieste University, Via G.B. Tiepolo, 11, 34131 Trieste \email{spitoni@oats.inaf.it}}
\maketitle

\abstract{We review chemical evolution models developed for M31 as well as the abundance determinations available for this galaxy. Then we present a recent chemical evolution model for M31 including radial gas flows and galactic fountains along the disk, as well as a model for the bulge. Our models are predicting the evolution of the abundances of several chemical species such as H, He, C, N, O, Ne, Mg, Si, S, Ca and Fe. From comparison between model predictions and observations we can derive some constraints on the evolution of the disk and the bulge of M31. We reach the conclusions that Andromeda must have evolved faster than the Milky Way and inside-out, and that its bulge formed much faster than the disk on a timescale $\le$0.5 Gyr. Finally, we present a study where we apply the model developed for the disk of M31 in order to study the probability of finding galactic habitable zones in this galaxy.}

\section{Introduction}
\label{sec:1}
The Andromeda galaxy is the spiral galaxy closest to the Milky Way (MW) and it is $\sim$1.5 times more massive. In early studies, it was pointed out that the average metallicity of M31 stellar halo ($<[M/H]> \sim$ -0.5 dex, e.g. Holland et al. 1996) is much higher than the average metallicity in the MW halo ($<[Fe/H]>\sim$ -1.8 dex, e.g. Ryan \& Norris, 1991). 
However, this conclusion is not fair since the comparison was made between the global metal content, M or Z, which is dominated by $\alpha$-elements, and Fe which evolves quite differently (see Matteucci, 2012). In particular, M31 halo stars could show an overabundance of $\alpha$-elements relative to Fe, as the Galactic halo stars. Therefore, if we assume [$\alpha$/Fe]=+0.3 dex for the stars in the M31 halo, we obtain that the $<[M/H]> \sim-0.5$dex transforms into ($<[Fe/H]>\sim$ -1.5dex, more similar to the metallicity of Galactic halo stars. In fact, Kalirai et al. (2006), by assuming  [$\alpha$/Fe]=+0.3 dex, derived a $<[Fe/H]>\sim$ -1.5dex for red giant branch stars in the outer halo of Andromeda. 
It should be reminded  that the chemical abundances of M31 stars are generally measured by means of color-magnitude diagrams, which are interpreted through stellar tracks computed for a given stellar metallicity Z (M). We should 
therefore be careful not to confuse Z with Fe, as it is still done in several papers. More recently, Koch et al. (2008) measured the [Fe/H] in the halo of M31 by means of spectroscopy (CaII triplet) and also found stars in the outermost region with [Fe/H] $<-2.0$ dex, thus implying that there is indeed a metal poor halo also in M31. 
Sarajedini \& Jablonka (2005) measured the metallicity distribution function (MDF) of stars in the bulge of M31 by means of the colour -magnitude diagram, and Worthey et al. (2005) also performed imaging of 11 fields in M31 with HST, and the chosen fields sample all galactocentric radii up to 50 kpc. The metallicity distributions thus obtained show a mild negative gradient that flattens outside 20 kpc along the M31 disk. An interesting fact is that stars in the outer regions of M31 are 10 times more metal rich than the MW halo stars and the globular clusters at the same galactocentric distance. This fact was interpreted as if the disk dominates over the halo at all radii out to 50 kpc. This can mean that many studies claiming to observe the halo of M31 were in reality observing the disk and this could be an additional fact, beyond the $\alpha$ elements versus Fe 
argument, to explain the apparent difference found between the average metallicity in the halo of M31 and that of the MW. 

Chemical abundances in M31 have been measured also 
from HII regions, planetary nebulae and young massive stars. In particular, 
Zurita \& Bresolin (2012), Sanders et al. (2012), Esteban et al. (2009) and Galarza et al. (1999) measured abundances in HII regions, while Przybilla et al. (2008), 
Trundle et al. (2002) and Venn et al. (2000) measured abundances in supergiants. On the other hand, Blair et al. (1982) and Dennefeld  \& Kunth (1981) derived abundances also from supernova remnants.

 The star formation rate in M31 has been estimated by Braun et al. (2009), Boissier et al. (2007) and Williams (2003a,b).

Chemical evolution models for M31 have been developed by Diaz \& Tosi (1984), Moll\'a et al. (1996), Renda et al. (2005), Ballero et al. (2007), Yin et al. (2009), Marcon-Uchida et al. (2010), Spitoni et al. (2013) and Robles-Valdez et al. (2013).  In most of these papers it has been concluded that M31 must have evolved faster than the MW with a higher efficiency of star formation. This can be expressed as a down-sizing in star formation, in analogy to what happens in elliptical galaxies (see for example Matteucci, 2012), since M31 is more massive than our Galaxy. Most of the models suggested also a star formation efficiency decreasing with increasing galactocentric distance in order to reproduce the O abundance gradient, except Spitoni et al. (2013) who included radial gas flows in their chemical model and did not find necessary to assume a variable star formation efficiency. It is well known, in fact, that radial gas flows have the effect of steepening the gradients (see Mott et al. 2013).

Here, we will review the main results of some of these papers and their conclusions and try to infer the history of formation and evolution of M31. Then we will present a study aimed at exploiting a detailed chemical evolution model for M31 in order to find the galactic habitable zones (GHZ) in this galaxy (Spitoni et al. 2014).

\section{Modelling chemical evolution of M31}
\label{subsec:2}
As suggested by Renda et al. (2005), when compared to the MW, Andromeda seems to have been more active in forming stars in the past although its current star formation rate (SFR) appears lower, since on average it has been estimated to be $1M_{\odot}yr^{-1}$ (Williams 2003a,b) against an average of 2-6$M_{\odot}yr^{-1}$ for the solar neighbourhood (Boissier \& Prantzos, 1999), which also can represents an average value along the disk. This means that M31 evolved faster than the MW: in other words, it must have had a higher star formation efficiency and/or a shorter timescale for gas accretion relative to the MW.
%This fact together with the larger mass of M31
%($(10.7-14)\cdot 10^{11}M_{\odot}$) relative to the Milky Way ($(4.0-5.5)\cdot M_{\odot}$, see Yin et al. 209 and references therein) suggests that there is a downs%izing in star formation also in spiral galaxies in the sense that more massive %spirals experiment a more efficient star formation than lower mass spirals, in analogy to what happens in elliptical galaxies. This means that Andromeda must have had a higher efficiency of star formation and/or a shorter timescale for gas accrteion than the Milky way.

The SFR, in most of the available models, is expressed by means of the Kennicutt law as :
\begin{equation}
SFR(r,t)= \nu \sigma_{gas}^{k}\,\, (M_{\odot}Gyr^{-1}pc^{-2})
\end{equation}
where $\sigma_{gas}$ is the surface gas density and $\nu$ is the SFR per unit mass of gas, namely the star formation efficiency, expressed in $Gyr^{-1}$. Kennicutt (1998) suggested an average $\nu=(0.25\pm0.07)Gyr^{-1}$ and $k=1.4\pm0.15$ for a sample of star forming galaxies.
Models of galactic chemical evolution adopting this formulation have adopted similar values for $\nu$ and $k$, always tuned to reproduce the observed properties of the studied galaxies.

Most chemical evolution models (Renda et al. 2005; Yin et al. 2009; Marcon-Uchida et al. 2010; Spitoni et al. 2013) assumed that the disk of Andromeda formed, like that of the MW, by accretion of cold gas with an infall law of the form:

\begin{equation}
I(r,t)= A(r)e^{-t/\tau(r)} (M_{\odot}Gyr^{-1}pc^{-2}),
\end{equation}
where A(r) is a parameter obtained by imposing:
\begin{equation}
\sigma_{tot}(r, T_{G})=\int_0^{T_{G}}{A(r)e^{-t/\tau(r)} dt}
\end{equation}
where $\sigma_{tot}(r, T_{G})$ is the present time surface mass density profile
and $\tau(r)$ is the time scale for gas accretion, which can be a function of galactocentric distance. In MW models it has been suggested (Matteucci \& Fran\c cois, 1989; Boissier \& Prantzos, 1999) that the disk forms inside-out and therefore it has been assumed that $\tau(r)$ is an increasing function of the galactocentric distance.
The present time surface mass density is an observed quantity and either in the MW or M31 it is well described by an exponential profile with a typical scale length.
The initial mass function is generally assumed to be constant in space and time and its form is either that of Scalo (1986) or Kroupa et al. (1993). 

Detailed chemical evolution models relax the instantaneous recycling approximation and consider the stellar lifetimes. In this way, stellar nucleosynthesis and therefore the contributions of supernovae (SNe) of different type can be taken into account precisely. With this kind of models one can predict the evolution in space and time of the abundances of several single chemical elements in the gas.
\begin{figure}[b]
\sidecaption
\includegraphics[width=8cm,angle=-90]{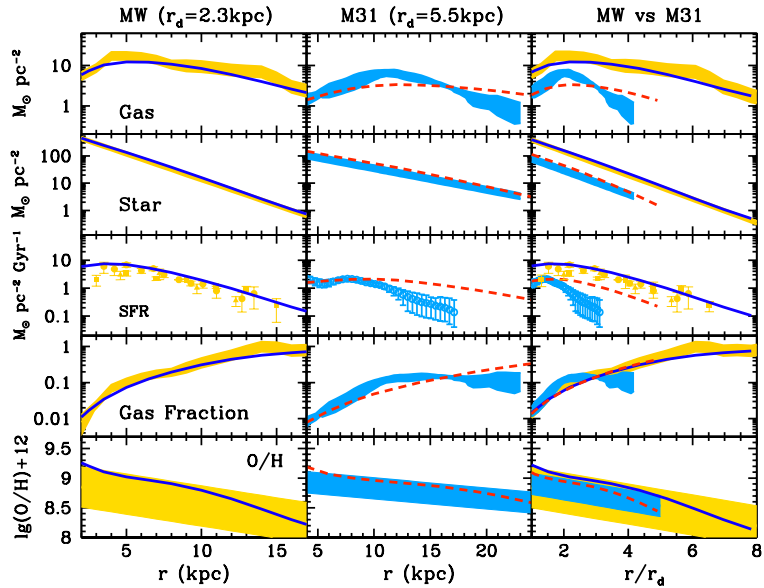}
\caption{MW versus M31: the observations are indicated by the shaded areas whereas the continuous line represents the model for the MW and the dashed line the model for M31. Figure 5 from Yin et al. (2009).}
\label{fig:1}       % Give a unique label
\end{figure}

In summary, once an IMF has been fixed, the main free parameters are the timescale of gas accretion, the efficiency of star formation and the exponent of the surface gas density in the Kennicutt law. In the following, we will present some results of chemical evolution models obtained by varying these three main parameters in order to fit the observed features of M31.

\subsection{Results about the chemical evolution of M31}

First we show the results of Yin et al. (2009) who adopted an inside-out formation for the disk of M31, a Kroupa et al. (1993) IMF and a Kennicutt law for the SFR with $k=1.5$ and $\nu_{M31}=0.2Gyr^{-1}$. They adopted similar assumptions for the MW model with the exception of the efficiency of star formation assumed to be $\nu_{MW}=0.1 Gyr^{-1}$, and the masses of the two disks which are $7.0\cdot10^{10}M_{\odot}$ for Andromeda and $5.0\cdot10^{10}M_{\odot}$ for the Milky Way. 
\begin{figure}[b]
\sidecaption
\includegraphics[width=10cm]{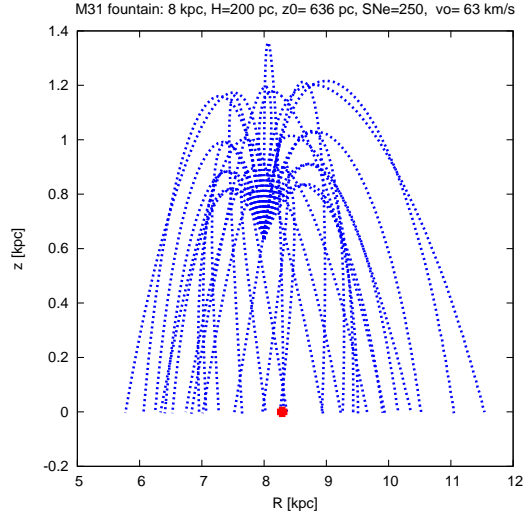}
\caption{Predicted galactic fountains in M31. The starting radius is 8 kpc and the height reached by the gas fountain is 636pc. The red filled circle indicates the average falling radius coordinate. Figure from Spitoni et al. (2013).}
\label{fig:1}       % Give a unique label
\end{figure}
\begin{figure}[b]
\sidecaption
\includegraphics[width=10cm]{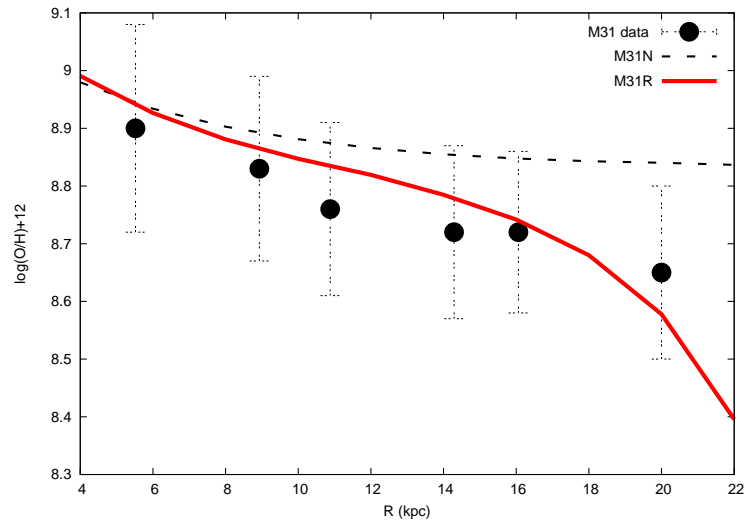}
\caption{Oxygen gradient in M31. The dotted line represents a model without radial gas flows, whereas the dashed line represents a model with gas flows (Spitoni et al. 2013). Data are from HII regions, young stars and SN remnants (Zurita \& Bresolin, 2012; Sanders et al.; 2012; Venn et al.; 2000; Esteban et al.; 2009;Galarza et al., 1999; Przybilla et al., 2008; Trundle et al., 2002; Blair et al., 1982; Dennefeld \& Kunth, 1981).To better understand the trend in the data, we divided them into six bins as functions of the galactocentric distance. In each bin, we computed the mean value and the standard deviation
for the oxygen abundance and the associated error.}
\label{fig:1}       % Give a unique label
\end{figure}
In Figure 1 we show the results of that paper which illustrate the differences between MW and M31. In particular, in this Figure are shown the predicted 
and observed behaviours of the gas, the stars, the SFR, the gas fraction and the O gradient ($\Delta(log(O/H)+12)$) along the disks of M31 and the MW. 

The model of Marcon-Uchida et al. (2010) assumed that the disk of M31 forms inside-out  with the time scale for gas accretion varying as:
\begin{equation}
\tau(r)=0.62r+1.62  (Gyr).
\end{equation}
The efficiency of star formation was also assumed to vary with r as:
\begin{equation}
\nu(r)=24/r-1.5  (Gyr^{-1})
\end{equation}
until it reaches a minimum value of $0.5Gyr^{-1}$, which is  kept constant for the largest radii. A threshold gas density for star formation of $5M_{\odot}pc^{-2}$ (Braun et al. 2009) was also adopted. The surface mass density profile for the disk was taken from Geehan et al. (2006) with a scale radius $R_D=5.4$ kpc, while in the MW $R_D=2.3-2.5$ kpc. Also this model predicts a faster evolution for Andromeda relative to the MW. In fact, although both disks are assumed to have formed inside-out, the time scales for gas accretion are shorter and the efficiencies of star formation are higher in Andromeda at the same galactocentric distance than what is generally assumed for the MW (e.g. Chiappini et al. 2001).

Spitoni et al. (2013) adopted a model similar to that of Marcon-Uchida et al. (2010) and included galactic fountains and radial gas flows with the aim of studying the effects of these processes on the O gradient in M31. By means of a ballistic method for the galactic fountains, they found that the landing coordinate of the gas expelled above the galactic disk from SNe is no more than 1 kpc from the starting point, thus not affecting the abundance gradients along the M31 disk (see Figure 2). They also found that including radial gas flows allowed them to reproduce very well the observed abundance gradient of oxygen without the necessity of assuming an efficiency of star formation strongly decreasing with radius or a threshold in the gas density for star formation. The existence of such a star formation threshold has, in fact, been questioned by GALEX observations (see Boissier et al. 2007). In Figure 3 we show the predicted O gradient for M31 compared with a large compilation of data including HII regions, young stars and SN remnants. The presence of radial gas flows derives from the assumption of the disk forming by infall of gas; in fact, the infalling gas has a lower angular momentum than that of the gas in the disk and the mixing of these two gases induces a net radial gas inflow.
Chemical evolution models with radial gas flows have been discussed by several authors in the past, including Lacey \& Fall (1985), Portinari \& Chiosi (2000), Sch\"onrich \& Binney (2009), Spitoni \& Matteucci (2011), but only Spitoni et al. (2013) applied the radial flow model to M31: in particular, they assumed a speed for the gas flow which is linearly increasing with decreasing galactocentric distance and it varies from 1.5 km/sec at 22 kpc down to 0.55km/sec at 2kpc.
\begin{figure}[b]
\sidecaption
\includegraphics[width=9cm]{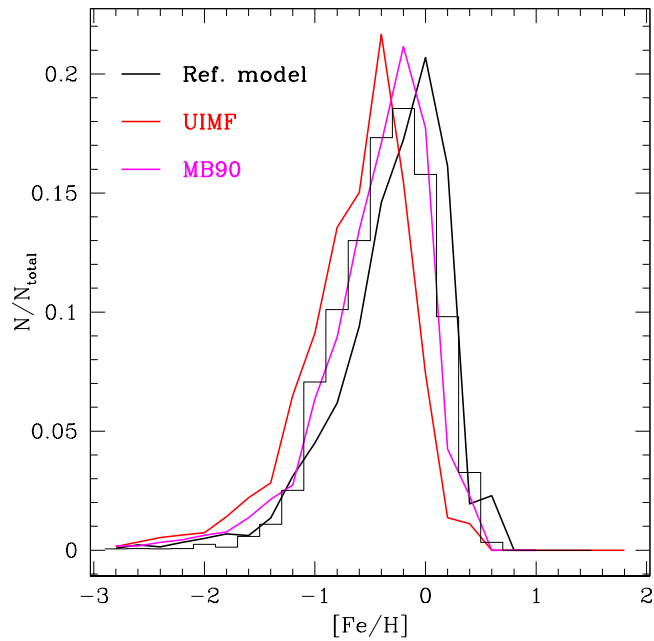}
\caption{MDF for the M31 bulge. The best fit for the M31 bulge is obtained with an IMF with x=1.1, which is indicated by MB90. The model labelled UIMF represents the universal IMF of Kroupa (2001). Figure from Ballero et al. (2007).}
\label{fig:1}       % Give a unique label
\end{figure}
\section{The chemical evolution of M31 bulge}
The chemical evolution of the bulge of M31, a classical bulge more massive than the bulge of the Milky Way, has been computed by Sarajedini \& Jablonka (2005), Worthey et al. (2005) and Ballero et al. (2007).
The first two models were simple models of chemical evolution with instantaneous recycling approximation, whereas the model of Ballero et al. (2007) was taking into account fast gas accretion, stellar lifetimes, detailed nucleosynthesis and SN progenitors. These last authors concluded that the bulge of M31 must have evolved very quickly and on a timescale no longer than 0.5 Gyr, with an intense burst of star formation with star formation efficiency of $\sim 20Gyr^{-1}$. This agrees with more recent studies (Saglia et al. 2010) who concluded that with the exception of the region in the inner arcsecs, the stars in the bulge of M31 are uniformly old ($\ge$ 12 Gyr) and with an overabundance of $\alpha$-elements
[$\alpha$/Fe]$\sim$ +0.2 dex and solar metallicity.  Sarajedini \& Jablonka (2005) derived the stellar metallicity distribution function (MDF) for M31, showing a peak at around the solar metal content (Z). In particular, the peak of the MDF of M31 bulge is at a metallicity higher by 0.1 dex relative to the MDF of the bulge of the MW.
In Figure 4 we report the predicted and observed MDFs for the M31 bulge; the predictions are from Ballero et al. (2007), who claimed that a flatter IMF than Salpeter (1955) and universal IMF of Kroupa (2001) is necessary to fit the data. In this Figure, the MDF is a function of [Fe/H]: the authors transformed the data from [M/H] to [Fe/H], by assuming $[\alpha/Fe]=+0.3$ dex as suggested in Sarajedini \& Jablonka (2005). All of these findings suggest that the bulge of M31 must have formed mainly by dissipational collapse during a strong burst of star formation.
\begin{figure}[b]
\sidecaption
\includegraphics[width=8cm,angle=-90]{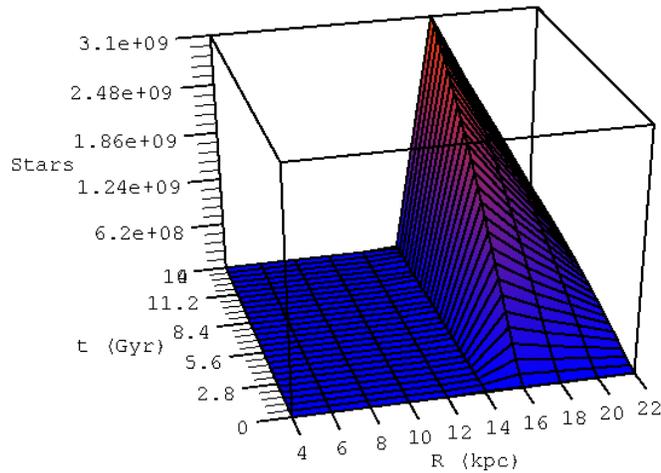}
\caption{A 3-D plot relative to the number of stars with habitable planets expected in M31. The chemical evolution model is that of Spitoni et al. (2013) with radial gas flows. Figure from Spitoni et al. (2014).}
\label{fig:1}       % Give a unique label
\end{figure}
\section{The GHZ in M31}
Finally, we show some predictions concerning the Galactic Habitable Zone (GHZ) for the disk of M31 (Spitoni et al. 2014).  We have assumed that the probability of forming Earth-like planets depends on the [Fe/H], the SFR and the SN rate of the studied region.
In particular, we define the effective probability of finding stars with Earth-like planets that survived SN explosions as a function of the galactic radius as:

\begin{equation}
P_{GHZ}={\int_0^R{SFR(r,t')P_E(r,t')P_{SN}(r,t')dt'} \over \int_0^R{SFR(r,t')dt'}},
\end{equation}
where $P_E(r,t)$ is the probability for the creation of Earth like planets and not giant planets and is a function of the Fe abundance, $P_{SN}(r,t)$ is the probability that a SN occurs close to the Earth-like planets and SFR(r,t) is the star formation rate. The metallicity, the star formation rate and the SN rate are those predicted by the Spitoni et al. (2013) model for M31. 

Therefore, the number of stars hosting Earth-like planets with possible life is given by the product of the probability of having such planets, $P_{GHZ}$, multiplied by the total numer of stars formed at each radius.
Figure 5 shows the predicted number of stars with habitable planets in Andromeda and the highest probability of finding them lies at 16 kpc from the center.

\section{Conclusions}
The Andromeda galaxy has an average metallicity higher than the MW but a shallower metal gradient along the disk at the present time, its bulge looks more evolved than the bulge of the MW, since it is mainly made of old stars exhibiting overabundances of $\alpha$-elements relative to Fe. All of this indicates that M31 has suffered a more intense star formation and and shorter timescales for the formation of all its components, than our Galaxy. On the other hand, M31 is more massive than the MW: this fact suggests that the baryonic galactic mass (stars plus gas) is linked to the star formation rate and that a down-sizing in star formation is present also in spiral galaxies. This process can be described by simply assuming that the efficiency of star formation is a function of the galactic baryonic mass. 
We also estimated which galactocentric region of the Andromeda disk has the highest probability of having stars hosting habitable planets and found that this region is centered at 16 kpc from the centre.
Future detailed abundance determinations will greatly help in assessing all the above mentioned points.

\begin{acknowledgement}
We aknowledge financial support from PRIN-MIUR2010-2011, project ``Chemical and Dynamical Evolution of the Milky Way and Local Group Galaxies'', prot. 2010LY5N2T.
\end{acknowledgement}
%
%%%%%%%%%%%%%%%%%%%%%%%% referenc.tex %%%%%%%%%%%%%%%%%%%%%%%%%%%%%%
% sample references
% %
% Use this file as a template for your own input.
%
%%%%%%%%%%%%%%%%%%%%%%%% Springer-Verlag %%%%%%%%%%%%%%%%%%%%%%%%%%
%
% BibTeX users please use
% \bibliographystyle{}
% \bibliography{}

\begin{thebibliography}{99.}%
\bibitem{}Ballero, S., Kroupa, P. Matteucci, F. 2007, A\&A, 467, 117
\bibitem{}Blair, W. P., Kirshner, R. P., Chevalier, R. A. 1982, ApJ, 254, 50
\bibitem{}Boissier S. \& Prantzos, N., 1999, MNRAS, 307, 857
\bibitem{}Boissier et al., 2007, ApJS, 173, 524
\bibitem{}Braun, R., Thilker, D.A., Walterbos, R.A.M.,Corbelli, E., 2009, ApJ, 695, 937
%\bibitem{}Chiappini, C., Matteucci, F., Gratton, R., 1997, ApJ, 477, 765
\bibitem{}Chiappini, C., Matteucci, F., Romano, D., 2001, ApJ, 554, 1044
\bibitem{}Dennefeld, M.,  Kunth, D., 1981, AJ, 86, 989
\bibitem{}Diaz, A.\& Tosi, M., 1984, MNRAS, 208, 365
\bibitem{}Esteban C., Bresolin F., Peimbert M., García-Rojas J., Peimbert A., 
Mesa-Delgado A., 2009, ApJ, 700,654
\bibitem{}Galarza, V. C., Walterbos, R. A. M., Braun, R. 1999, AJ, 118, 2775
\bibitem{}Geehan, J.J., Fardal, M.A., Babul, A., Guhathakurta, P., 2006, MNRAS, 366, 996
%\bibitem{}Goetz, M.\& Koeppen, J., 1992. A\&A, 262, 455
\bibitem{}Holland, S., Fahlman, G.G., Richer, H.B., 1996, AJ, 1121035
\bibitem{}Kalirai, J.S. et al., 2006, ApJ, 648,389
\bibitem{}Kennicutt, R.C. Jr.,  1998, ARA\&A, 36, 189
\bibitem{}Koch, A. et al. 2008, ApJ, 689, 958
\bibitem{}Kroupa, P., 2001, MNRAS, 332, 231
\bibitem{}Kroupa, P., Tout, C.A., Gilmore, G., 1993, MNRAS, 262, 545
\bibitem{}Lacey, C.G.\& Fall, M., 1985, ApJ, 290, 154
\bibitem{}Marcon-Uchida, M., Matteucci, F., Costa, R., 2010, A\&A, 520, 35
\bibitem{}Matteucci. F., 2012, Chemical Evolution of Galaxies: Astronomy and Astrophysics Library. ISBN 978-3-642-22490-4. Springer-Verlag Berlin Heidelberg, 2012
\bibitem{}Matteucci. F. \& Fran\c cois, P., 1989, MNRAS, 239, 885
\bibitem{}Moll\'a, M., Ferrini, F., Diaz, A., 1996, ApJ, 466, 668
\bibitem{}Mott, A., Spitoni, E., Matteucci, F.,  2013, MNRAS, 435, 2018
\bibitem{}Portinari, L. \& Chiosi, C., 2000, A\&A, 355, 929
\bibitem{}Przybilla N, Butler K., Kudritzky R., 2008, in "The Metal Rich Universe" Conference Proceedings,  Cambridge University Press, Cambridge, U. K., 2008, p.332
\bibitem{}Renda, A., Kawata, D., Fenner, Y., Gibson, B.K., 2005, MNRAS, 356, 1071
\bibitem{}Robles-Valdez, F., Carigi, L., Peimbert, M., 2013, arXiv:1310.1420
\bibitem{}Ryan, S.G. \& Norris, J.E., 1991, AJ, 101, 1865
\bibitem{}Saglia, R.P. et al., 2010, A\&A, 509, 61 
\bibitem{}Sanders N., Caldwell N., McDowell J., Harding P., 2012, ApJ, 758, 
133S
\bibitem{}Sarajedini, A. \& Jablonka, P., 2005, ApJ, 130, 1627
\bibitem{}Salpeter, E.E., 1955, ApJ, 121, 161
\bibitem{}Scalo, J.M., 1986, FCPh, 11, 1
\bibitem{}Schoenrich, R., Binney, J.,  2009, MNRAS, 396, 203
\bibitem{}Spitoni, E. \& Matteucci,F., 2011, A\&A, 531, 72
\bibitem{}Spitoni, E., Matteucci,F.,  Marcon-Uchida,M.,  2013, A\&A, 551, 123
\bibitem{}Spitoni,E., Matteucci, F.,  Sozzetti, A., 2014, MNRAS, 440, 2588 
\bibitem{}Trundle C., Dufton P., Lennon D., Smartt S., Urbaneja M., 2002, A\&A, 
395, 519
\bibitem{}Venn K., McCarthy J., Lennon D., Przybilla N., Kudritzki R., 
Lemke M., 2000, ApJ, 541, 610
\bibitem{}Williams, B.F., 2003a, AJ, 126, 1312
\bibitem{}Williams, B.F., 2003b, MNRAS, 340, 143
\bibitem{} Worthey, G., Espa{\~n}a, A., MacArthur, L.~A., \& Courteau, S., 2005, ApJ, 631, 820 
\bibitem{}Yin, J., Hou, J.~L., Prantzos, N., et al., 2009, A\&A, 505, 497 
\bibitem{}Zurita A., Bresolin F., 2012, MNRAS, 427, 1463
\end{thebibliography}
%

\end{document}